\documentclass[a4paper,11pt]{article}
\usepackage{pos}

\usepackage{caption}
\usepackage{subcaption}

\usepackage{definitions}
\usepackage{xspace}
\usepackage{booktabs}
\usepackage{graphics}
\usepackage{adjustbox}

\usepackage{natbib}

\title{
Higgs self-coupling measurements at the FCC-hh} 
\author*[a]{Birgit Stapf}
\author[c]{Angela Taliercio}
\author[a, b]{Elisabetta Gallo}
\author[a, b]{Kerstin Tackmann}
\author[d]{Paola Mastrapasqua}

\affiliation[a]{Institut für Experimentalphysik, Universität Hamburg, \\
                Luruper Chaussee 149, 22761 Hamburg, Germany}

\affiliation[b]{Deutsches Elektronen-Synchrotron DESY \\
                Notkestr. 85, 22607 Hamburg, Germany}

\affiliation[c]{Northwestern University, \\
                Evanston, Illinois, USA}

\affiliation[d]{ Centre for Cosmology, Particle Physics and Phenomenology (CP3), \\
                Université catholique de Louvain, \\ 
                Louvain-la-Neuve, Belgium}

\emailAdd{birgit.stapf@desy.de}

\abstract{
The hadron collider phase of the Future Circular Collider (FCC-hh) is a proton-proton collider operating at a center-of-mass energy of 100 TeV. 
It is one of the most ambitious projects planned for the rest of this century and offers ample opportunities in the hunt for new physics, both through its direct detection reach as well as through indirect evidence from precision measurements. 
Extracting a precision measurement of the Higgs self-coupling from the Higgs pair production cross-section will play a key role in our understanding of electroweak symmetry breaking, as the self-coupling gives insight into the nature of the Higgs potential. 
With the large data set of in total 30 $\text{ab}^{-1}$ which is envisioned to be collected during the FCC-hh runtime the Higgs self-coupling will be determined down to the percent level. 
This paper presents prospect studies for Higgs self-coupling measurements in the \bbyy and \bbllMET final states, with the combined, expected precision on the Higgs self-coupling modifier \kappaLambda reaching 3.2-5.7\% at 68\% confidence level, assuming all other Higgs couplings follow their Standard Model expectations and depending on the systematic uncertainties assumed. This high precision is mostly driven by the \bbyy final state analysis, while the \bbllMET final state - newly studied for its FCC-hh prospects in this document - on its own reaches a maximum precision of roughly 20\% on \kappaLambda. 
}

\FullConference{%
  The European Physical Society Conference on High Energy Physic (EPS-HEP 2023)\\
  21-25 August 2023\\
  Hamburg, Germany
}

\begin{document}
\maketitle

\section{Introduction}
\noindent The study of Higgs boson pair production is one of the key benchmarks of the scientific program at future colliders. It offers direct experimental access to the Higgs boson trilinear self-coupling and hence to the structure of the scalar potential itself, allowing unprecedented insight into the electroweak symmetry breaking mechanism. 
The self-coupling modifier \kappaLambda, defined as the ratio of the measured value of the self-coupling over its Standard Model (SM) predicted value - $\lambda/\lambda_{SM}$ - is used to parameterise any deviation from the SM expectations. 
A precision measurement of \kappaLambda down to the percent level is a clear goal for future experiments, as at the end of the HL-LHC era the \kappaLambda precision is expected to reach only around 50\%~\cite{ATLAS_HL_LHC_HH}. 
%In the SM, Higgs pair production events in hadron collisions are a rare occurence - with the production cross-section for a Higgs pair in the dominant gluon-gluon fusion mode being three orders of magnitude smaller than that to produce a single Higgs boson~\cite{hh_theory}. 
The envisioned proton-proton collision phase of the Future Circular Collider (FCC-hh) would provide a total data set of \fcclumi, at a centre-of-mass energy of \sqrts = 100 TeV~\cite{CDR}, making a precision measurement of \kappaLambda possible. 
% as the HL-LHC dataset is expected to  \kappaLambda down to 
% constituting an expected improvement of a factor 20 in the precision on \kappaLambda, compared to the HL-LHC dataset~\cite{ATLAS_HL_LHC_HH}. 
%With such a large dataset, also a measurement of \kappaLambda in rarer or more complex final states of the Higgs pair becomes possible. 

\noindent In this document, we present an updated strategy for the analysis of the \bbyy final state with respect to previous self-coupling studies at FCC-hh~\cite{fcc_dihiggs_paper}, as well as a new study for FCC-hh using \bbllMET final state events. 

\section{Event generation, detector simulation and data analysis}
\noindent Proton-proton collision events at \sqrts = 100 TeV using a fast detector simulation with the \delphes framework~\cite{delphes} form the basis of the prospect studies. The signals, constituted by gluon-gluon fusion induced double Higgs production events for different values of \kappaLambda, are generated with \powheg ~\cite{powheg_one, powheg_two, powheg_three} at next-to-leading order (NLO). 
The respective cross-sections are scaled to NNLO using a $k$-factor that is independent of \kappaLambda ~\cite{hh_nnlo} (following~\cite{fcc_dihiggs_paper} and motivated by the studies in~\cite{gghh_nnlo_kfactor}). 
All background processes are generated with \madgraph ~\cite{madgraph_1, madgraph_2}. 
For all samples, the hadronisation effects as well as the Higgs decays are modeled with \pythia ~\cite{pythia}. 
%A fast detector simulation is applied, using the \delphes framework~\cite{delphes}. 
An optimistic, nearly ideal detector is assumed, with identification efficiencies reaching above 90\% and momentum resolutions in the per mille range, parameterised in terms of the kinematic properties of each object. %~\cite{delphes_card}.

%where the reconstruction and identification efficiencies as well as (momentum) resolutions are parametrized in terms of the kinematic properties of each object~\cite{delphes_card}. Here, photons, muons and electrons are typically identified with efficiencies around or above 90\%, with a relative momentum resolution below 1\%. The b-tagging efficiency is at least 80\% in the phase space considered by the analyses. 

\noindent  It is important to highlight that the technical implementation of all the above steps is part of the \keyhep project~\cite{key4hep}, that provides a consistent software stack for \textit{all} future collider facilities. In particular, the samples used here are in \edmhep format~\cite{edm4hep}. %, and produced with \ksimdelphes~\cite{k4simdelphes}. 
These samples are processed with the \FCCAnalyses framework~\cite{fccanalyses}.

\subsection{\texorpdfstring{\bbyy analysis}{bb̅γγ analysis}}
\noindent Although the \bbyy final state is rare, with a branching ratio of 0.26\%, it provides excellent sensitivity on \kappaLambda due to its clean signature with well-reconstructed objects, namely two photons and two b-jets. %with each pair having an invariant mass around the Higgs mass. 
The full $HH$ system can be reconstructed with good resolution. 
Backgrounds arise from single Higgs events and the non-resonant QCD-induced production of two isolated, energetic photons. 

\begin{figure}[h]
    \centering
    \begin{subfigure}[b]{0.67\textwidth}
        \includegraphics[width=\textwidth]{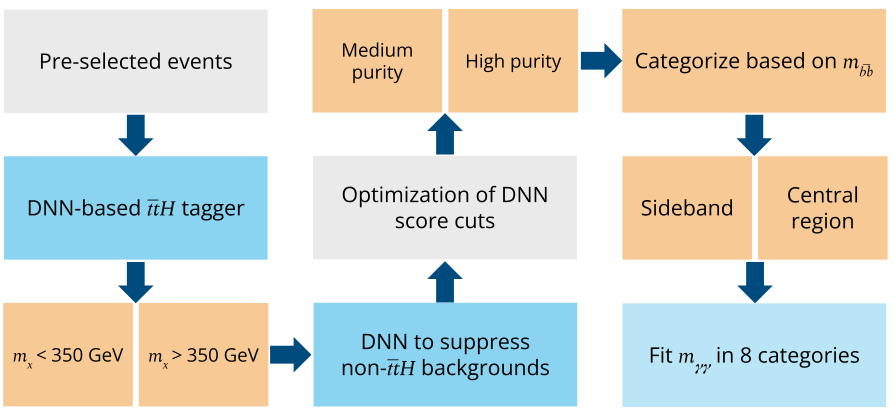}
        \caption{}
        \label{fig:bbyy_overview}
    \end{subfigure} 
    \hfill
    \begin{subfigure}[b]{0.32\textwidth}
        \includegraphics[width=\textwidth]{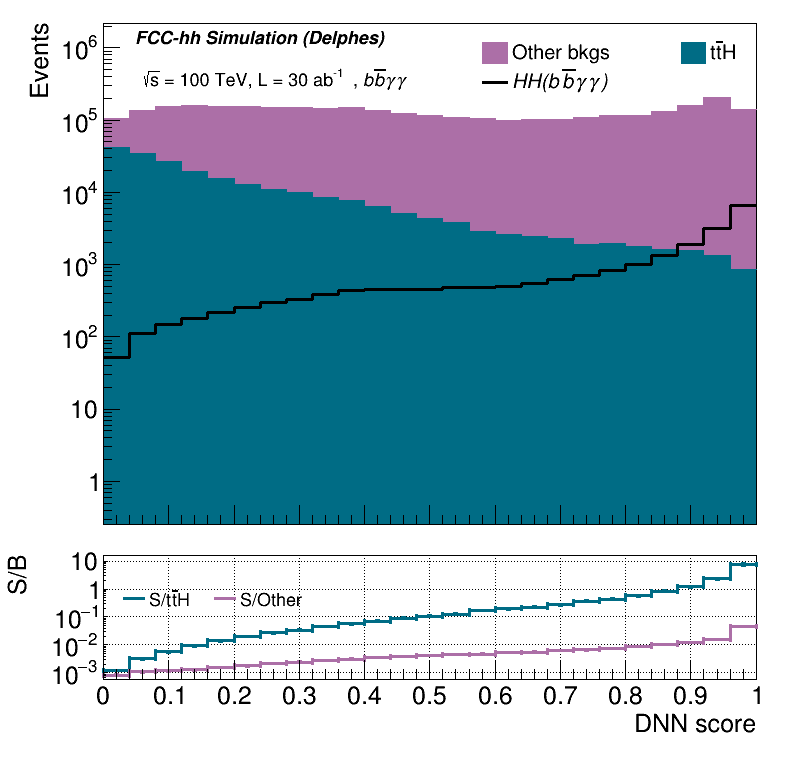}
        \caption{}
        \label{fig:bbyy_ttH_tagger} 
    \end{subfigure}\\
    \hfill
    \begin{subfigure}[b]{0.32\textwidth}
        \includegraphics[width=\textwidth]{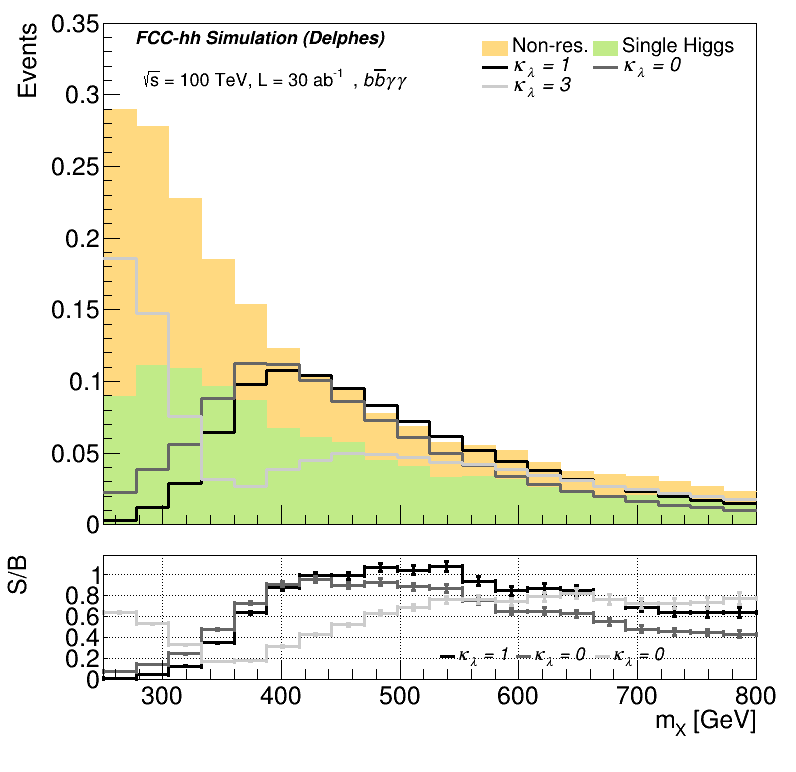}
        \caption{}
        \label{fig:bbyy_mx}
    \end{subfigure}
    \hfill
    \begin{subfigure}[b]{0.32\textwidth}
        \includegraphics[width=\textwidth]{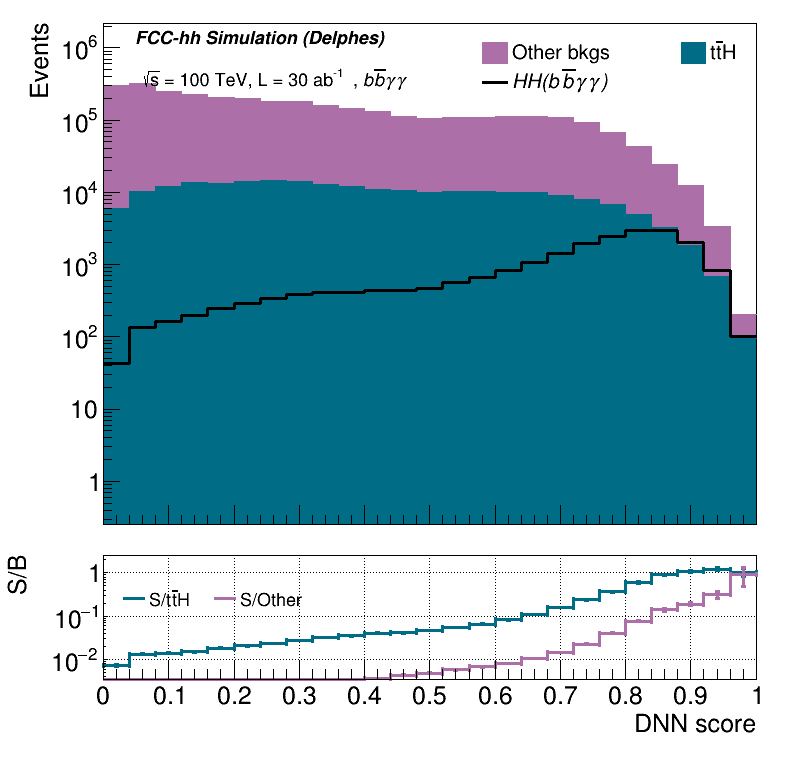}
        \caption{}
        \label{fig:bbyy_glob_DNN}
    \end{subfigure}
    \hfill
    %     \begin{subfigure}[b]{0.3\textwidth}
    %     \includegraphics[width=\textwidth]{./plots/plot_mbb_w_sideband_indication.png}
    %     \caption{}
    %     \label{fig:bbyy_mbb}
    % \end{subfigure}
    % \hfill
    \begin{subfigure}[b]{0.32\textwidth}
        \includegraphics[width=\textwidth]{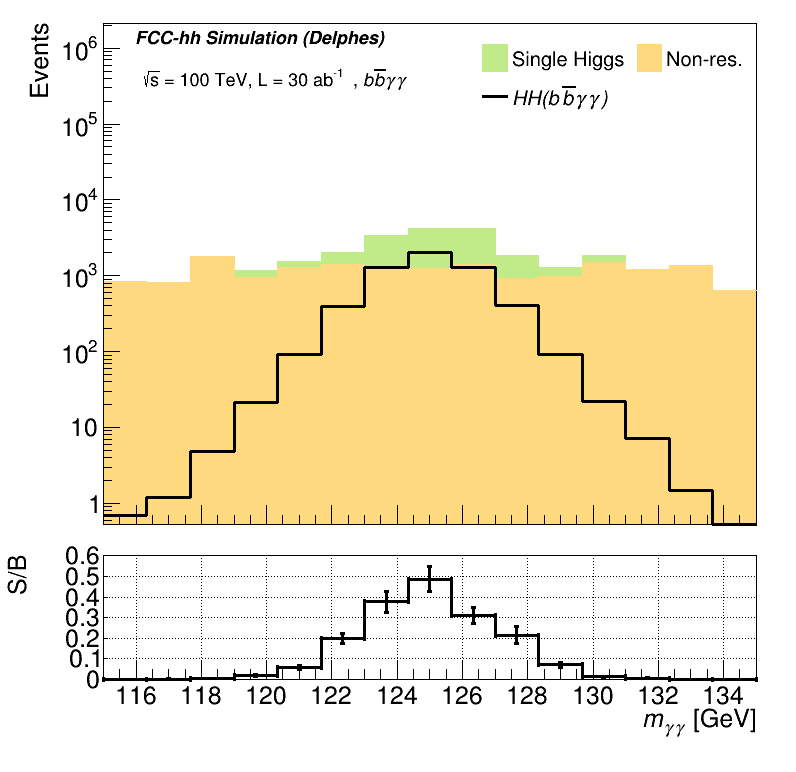}
        \caption{}
        \label{fig:bbyy_myy_central}
    \end{subfigure}
    \hfill
    % \begin{subfigure}[b]{0.3\textwidth}
    %     \includegraphics[width=\textwidth]{./plots/plot_myy_sideband.png}
    %     \caption{}
    %     \label{fig:bbyy_myy_sb}
    % \end{subfigure}
    \hfill
\caption{a) Overview of the \bbyy analysis strategy. 
b) Distributions of the DNN score of the \ttH tagger. 
c) Distributions of the reconstructed \mx for different anomalous coupling hypotheses and the background. 
d) Distributions of the DNN score to suppress non-\ttH backgrounds, for events with \mx $>$ 350 GeV. 
%e) Distributions of the invariant mass \mbb used to define central region and sidebands. 
%f) Distributions of the invariant mass \myy in the central region. 
%And g) distributions of the invariant mass \myy in the sidebands. 
And e) distributions of the invariant mass \myy in an example central region. 
}
\label{fig:bbyy}
\end{figure}

\noindent To maximize the sensitivity, a multi-variate analysis strategy is employed, relying on different Deep Neural Networks (DNN) to suppress the different backgrounds. The DNNs are implemented using a Keras frontend~\cite{keras} with a tensorflow backend~\cite{tensorflow}. Figure~\ref{fig:bbyy_overview} summarizes the various steps of the analysis strategy. 
%After a basic pre-selection requiring the expected two photons and b-jets, a first DNN is trained to distinguish only between the signals (with  different \kappaLambda) and the \ttH background. 
First, a DNN is trained to differentiate between the signals and the \ttH background.
The contribution of the \ttH background is enhanced among the single Higgs production modes since it results in a similar final state as the signal. However, its characteristic kinematics differ from those of the signal: 
generally, \ttH events have more, but less energetic jets and/or leptons with large transverse momentum, and in the signal the photon and b-jet pair are expected to be back-to-back, so with a large angle between them, but small angles within each pair.
% Generally, \ttH events tend to have more jets, that are less energetic. Additionally, \ttH events can contain charged leptons with high transverse momentum from the top-quark decays. 
% Finally, in the signal the photon and b-jet pair are expected to be back-to-back, so with a large angle between their decay systems, but small angles between themselves. 
All of this information is exploited by the DNN based \ttH-tagger. 
As can be seen in Figure~\ref{fig:bbyy_ttH_tagger} showing the DNN scores, the tagger provides good separation between signal and the targeted \ttH background. % while the remaining backgrounds are not so well separated. 
For the next step, events are divided into two categories based on the invariant mass $\mx = m_{b\bar{b}\gamma\gamma} - m_{b\bar{b}} - m_{\gamma\gamma} + 250 \, \text{GeV}$, which reconstructs the di-Higgs mass corrected for resolution effects. 
As illustrated in Figure~\ref{fig:bbyy_mx}, the shape of this distribution depends on \kappaLambda. 
Two separate DNNs, with the same setup and input variables as the \ttH tagger are trained to discriminate the signals from the remaining backgrounds. Figure~\ref{fig:bbyy_glob_DNN} shows the resulting scores for the example of events with \mx$>$ 350 GeV. 
A multi-dimensional optimization procedure with the significance as the figure of merit is performed to 1) discard events below a lower threshold on the DNN scores and 2) define a medium and high purity region based on the score of the second DNN. 
Last, events are categorized into a central or sideband region according to the value of the invariant mass of the b-jet pair, \mbb, using a window around the Higgs mass of around 15 GeV. 
%as illustrated in Figure~\ref{fig:bbyy_mbb}. 
This full procedure results in eight categories, in which the invariant diphoton mass \myy is used in the likelihood fit to extract \kappaLambda. 
Figure~\ref{fig:bbyy_myy_central} shows the \myy distributions in an example central region.
%Figures~\ref{fig:bbyy_myy_central} and~\ref{fig:bbyy_myy_sb} show the \myy distributions in an example central and sideband region. 

%test minipage 
% \begin{figure}[!htb]
%     \centering
%     \begin{minipage}{0.5\textwidth}
%         \centering
%         \includegraphics[width=\textwidth]{./plots/bbyy_strategy_flowchart.png}
%         \caption{$dt=0.1$}
%         \label{fig:prob1_6_2}
%     \end{minipage}
%     \\
%     \begin{minipage}{0.5\textwidth}
%         \centering
%         \includegraphics[width=\textwidth]{./plots/bbyy_kin_sketch-2.pdf}
%         \caption{$dt =$}
%         \label{fig:prob1_6_1}
%     \end{minipage}
%     \begin{minipage}{0.5\textwidth}
%         \centering
%         \includegraphics[width=\textwidth]{./plots/bbyy_H_bkg_comp.pdf}
%         \caption{$dt =$}
%         \label{fig:prob1_6_1}
%     \end{minipage}
% \end{figure}

\subsection{\texorpdfstring{\bbllMET analysis}{bb̅ℓℓ + ETMiss}}
\noindent The \bbllMET analysis considers the sum of signals from the \HHbbWWlvlv, \HHbbtautaulvlv and \HHbbllvv Higgs pair decays. Together, their branching ratio amounts to 3.24\%, so these type of events are an order of magnitude more common than \bbyy events. 
Nonetheless, the \bbllMET analysis is more difficult. 
The final state involves missing transverse energy \met from neutrinos escaping detection, so reconstructing one of the two Higgs decays fully is not possible. 
Moreover, the dileptonic decay of a top quark pair leads to the same the final state: two b-jets, two charged light leptons (electrons or muons) and \met. The production cross-section of this irreducible \ttbar background is seven orders of magnitude larger than that of the \bbllMET signal. Additional backgrounds arise from single Higgs and single top production, the \Vjets Drell-Yan process and the production of a top quark pair together with vector boson(s). 

\begin{figure}[h]
    \centering
    \begin{subfigure}[b]{0.32\textwidth}
        \includegraphics[width=\textwidth]{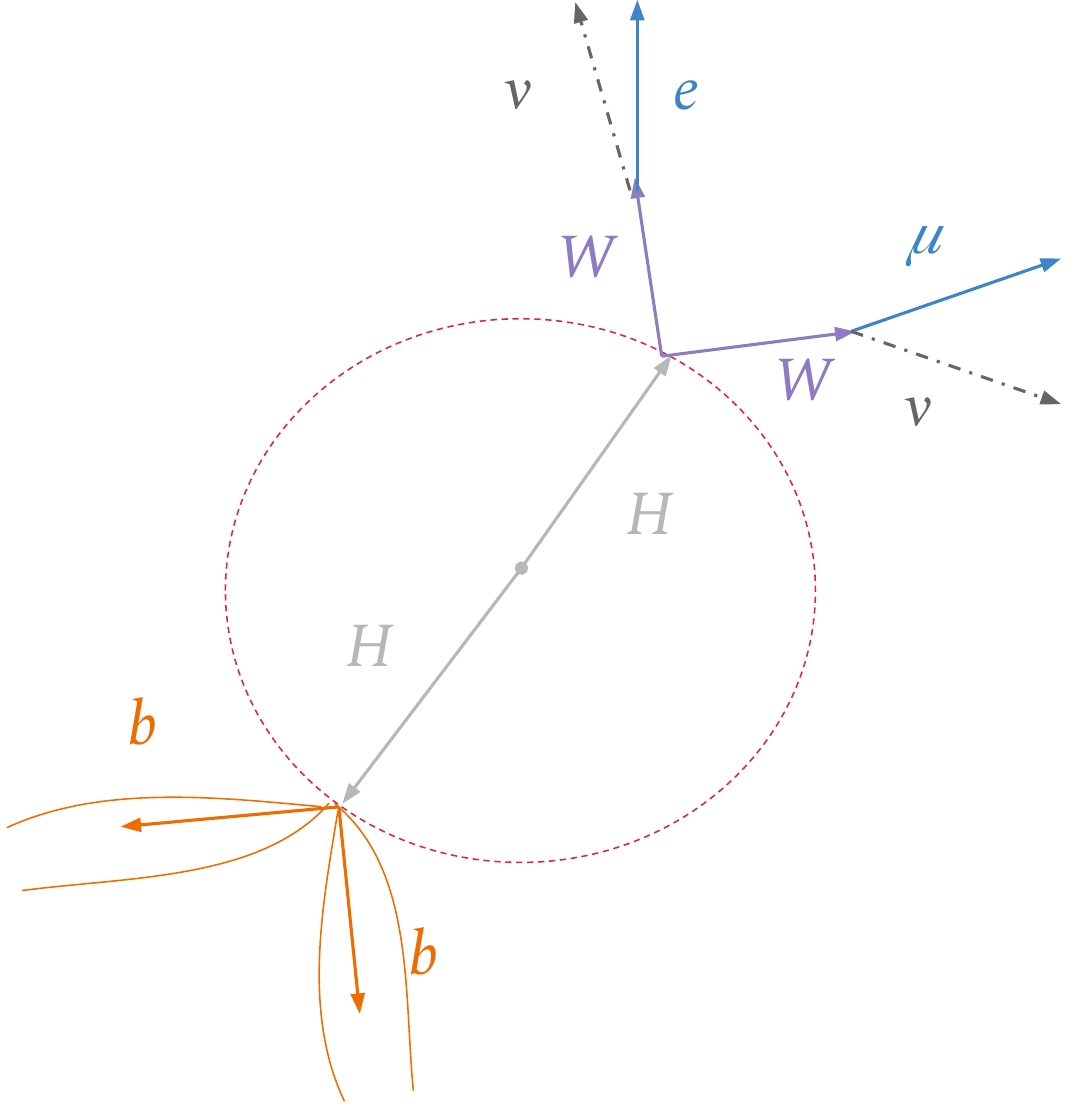}
        \caption{}
        \label{fig:bb2lmet_sketch}
    \end{subfigure}
    \hfill
    \begin{subfigure}[b]{0.32\textwidth}
        \includegraphics[width=\textwidth]{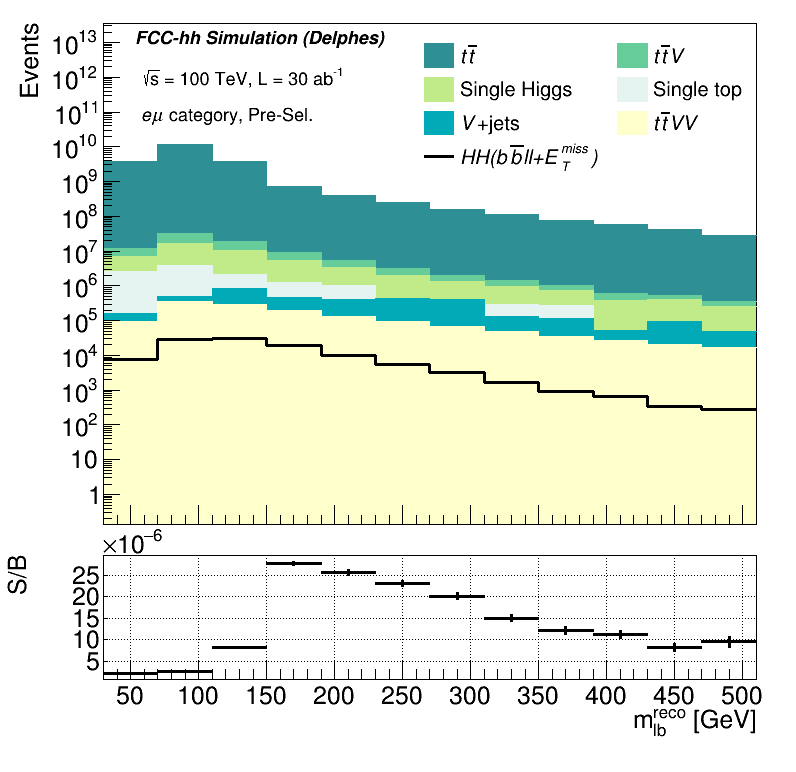}
        \caption{}
        \label{fig:bb2lmet_mlbreco}
    \end{subfigure}
    \hfill
    \begin{subfigure}[b]{0.32\textwidth}
        \includegraphics[width=\textwidth]{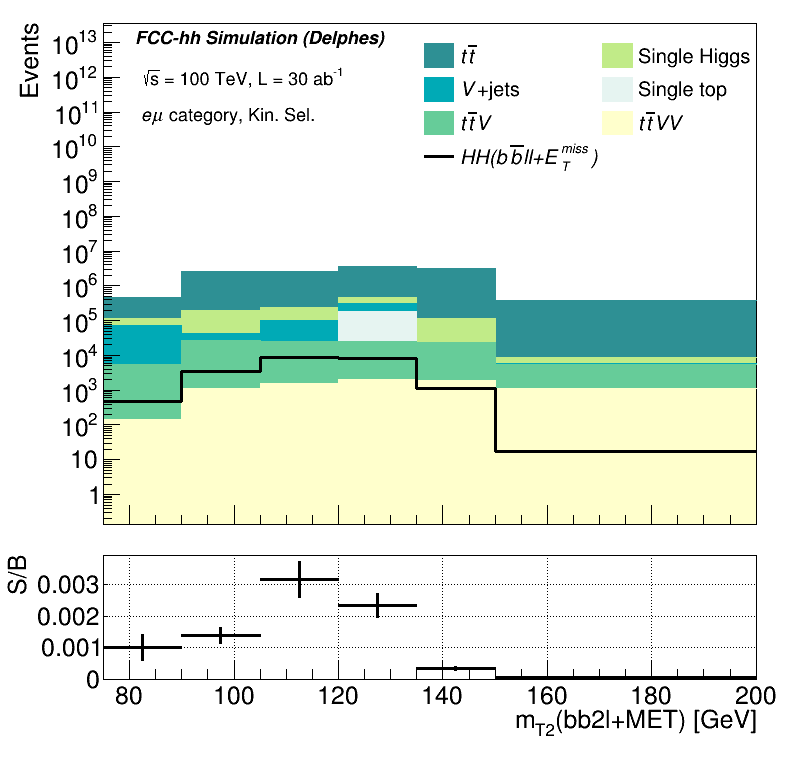}
        \caption{}
        \label{fig:bb2lmet_mt2}
    \end{subfigure}
\caption{a) Sketch of (idealized) \HHbbWWlvlv event kinematics. b) Distributions of the \mlb variable in events with an electron muon pair and two b-jets. And c) distributions of the stransverse mass \mstrans after full kinematic selection, in the electron-muon category.}
\label{fig:bb2lmet}
\end{figure}

\noindent A cut-based analysis is implemented to enhance sensitivity, exploiting the expected signal kinematics sketched in Figure~\ref{fig:bb2lmet_sketch}. In particular, the overwhelmingly large \ttbar background can be efficiently suppressed employing a lower bound on the minimum average invariant mass of the lepton and b-jet pairs, \mlb $=\text{min} \left(  \frac{m_{\ell_1b_1} + m_{\ell_2b_2}}{2}, \frac{m_{\ell_2b_1} + m_{\ell_1b_2}}{2}  \right) $, which is used in measurements of the top quark mass~\cite{ATLAS_top_mass_dilep}. Figure~\ref{fig:bb2lmet_mlbreco} shows the distributions of this variable in signal and background. To capture the full $HH$ decay, the \textit{stransverse mass} \mstrans \cite{stranverse_mass} which predicts the invisible mass contribution from the neutrinos is used in the likelihood fit in this channel. This is done in five categories based on the flavours of the leptons, whether a resonant $Z$-decay is present and the angle between the leptons and \met. Figure~\ref{fig:bb2lmet_mt2} shows the \mstrans distributions in an example category. 

\section{Results} 
\noindent The determination of the Higgs self-coupling modifier \kappaLambda is performed for three different scenarios of assumed systematic uncertainties, as listed in Table~\ref{tab:syst_scens}. Generally the systematic scenarios considered are optimistic (and in line with previous FCC-hh studies~\cite{fcc_dihiggs_paper}), assuming for example that for the \bbllMET analysis the large \ttbar and \Vjets backgrounds will be measured in data control regions with high precision. The uncertainties are only considered as impact on the event rates, not on the kinematic distributions of the process they apply to. 

\begin{table}[h]
    
    \begin{minipage}{0.7\linewidth}
    \begin{adjustbox}{width=\columnwidth,center}
        \begin{tabular}{l l c c c c }
            & \textbf{Source of uncertainty} & \textbf{Syst. 1} & \textbf{Syst. 2} & \textbf{Syst. 3} & \textbf{Applies to}   \\ 
            \toprule
            % \textbf{Common systematics} & &    &    & \\
            % \midrule
            & b-jet ID per b-jet & 0.5\% & 1\% & 2\% & Signals, MC bkgs.  \\
            \textit{Common} & Luminosity  & 0.5\%  & 1\% & 2\% & Signals, MC bkgs.  \\
            & Signal cross-section  & 0.5\%  & 1\% & 1.5\% & Signals, MC bkgs.   \\
            \midrule
            % \textbf{$b\bar{b}\gamma\gamma$ systematics} &    &    & \\
            % \midrule
            \text{\bbyy} & $\gamma$ ID per $\gamma$ & 0.5\% & 1\% & 2\% & Signals, MC bkgs.  \\
            \midrule
            % \textbf{\bbllMET systematics} &    &    & \\
            % \midrule
            & Lepton ID per lepton & 0.5\% & 1\% & 2\% & Signals, MC bkgs.  \\
            \textit{\bbllMET} & Data-driven bkg. est. & - & 1\% & 1\% &  \Vjets \\
            & Data-driven bkg. est. & - & - & 1\% &  \ttbar \\
            \bottomrule
        \end{tabular}
    \end{adjustbox}
    \caption{Overview of systematic uncertainties considered.  } 
    \label{tab:syst_scens}
    \end{minipage}
    \hfill
    \begin{minipage}{.25\linewidth}
    \begin{adjustbox}{width=0.95\columnwidth,center}
        \begin{tabular}{l  c }
        \textbf{Uncertainties} & \textbf{$\delta$\kappaLambda (68\% CL)} \\
        \toprule
        Stat. only & 3.2\% \\
        Syst. 1 & 3.6\% \\
        Syst. 2 & 3.9 \% \\
        Syst. 3 & 5.7 \% \\
        \bottomrule
        \end{tabular}
    \end{adjustbox}
    \caption{Expected precision on \kappaLambda when combining the \bbyy and \bbllMET analyses.} 
    \label{tab:syst_results}
    \end{minipage}

\end{table}

\begin{figure}[h]
    \begin{subfigure}[b]{0.49\textwidth}
        \centering
        \includegraphics[width=\textwidth]{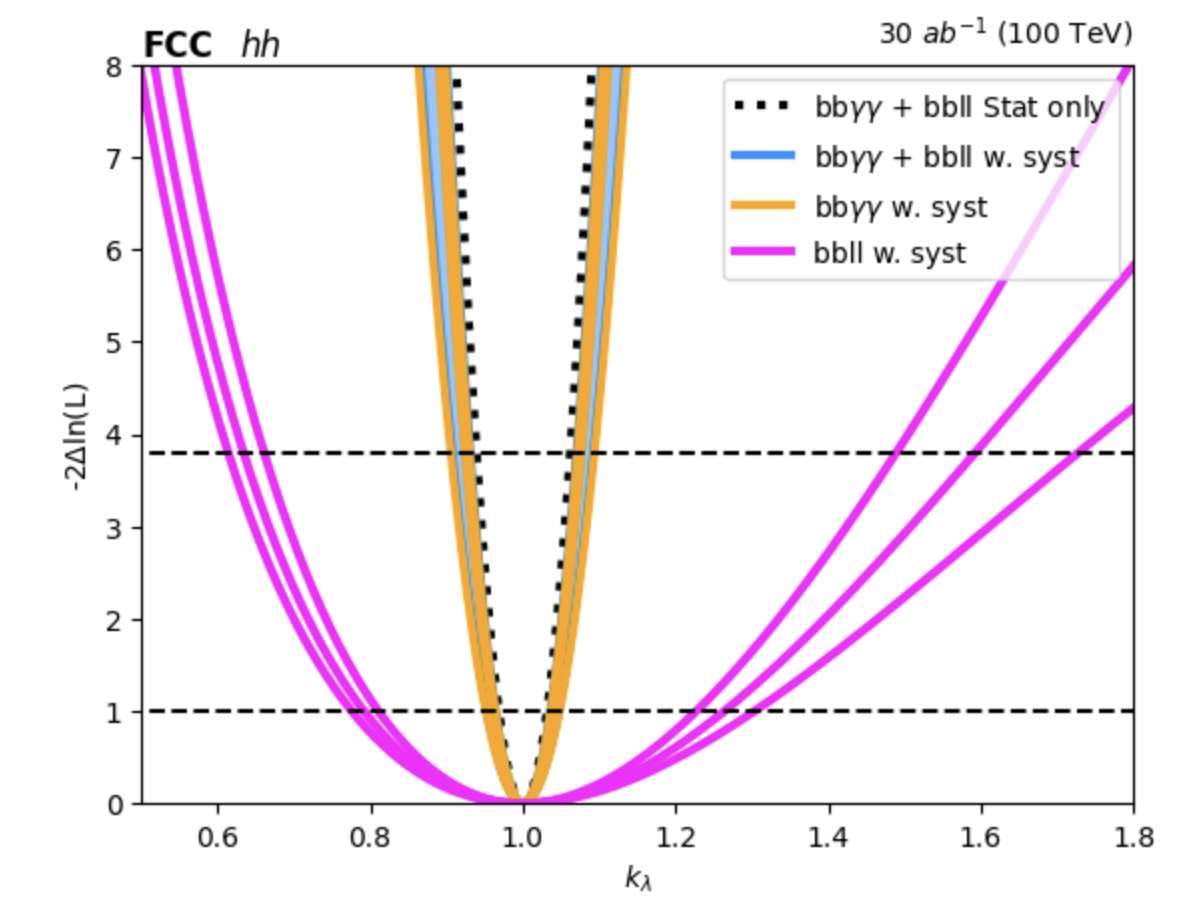}
        \caption{}
        \label{fig:result_kl_1D}
    \end{subfigure}
    \hfill
    \hfill
    \begin{subfigure}[b]{0.49\textwidth}
        \centering
        \includegraphics[width=0.8\textwidth]{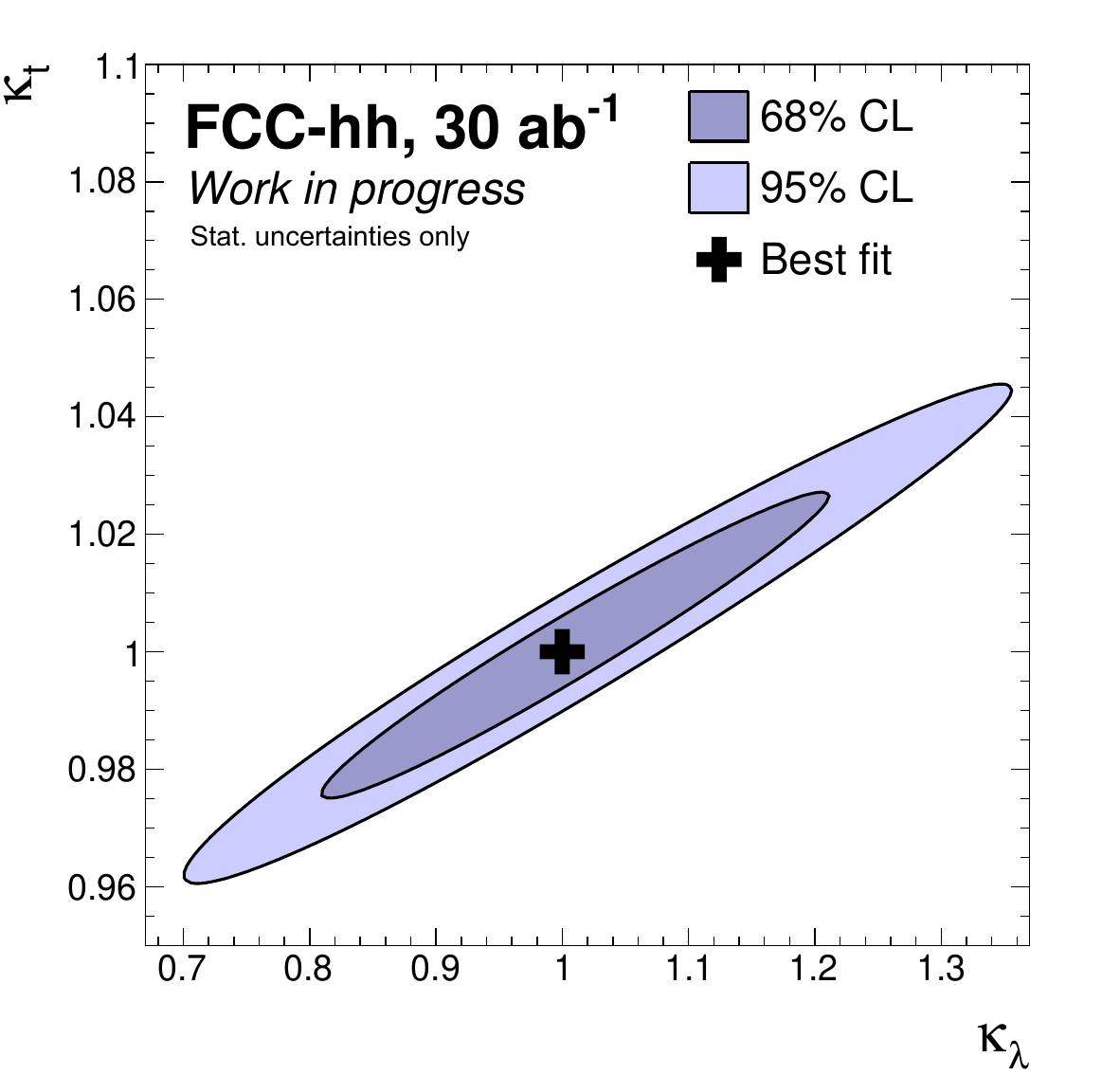}
        \caption{}
        \label{fig:result_kl_kt_2D}
    \end{subfigure}
\caption{a) Expected likelihood scans as a function of \kappaLambda for the \bbyy and \bbllMET analyses and their combination. b) Expected likelihood scan as a function of \kappaTop and \kappaLambda for the \bbyy and \bbllMET combination.}
\end{figure}

\noindent To extract the Higgs self-coupling modifier \kappaLambda from the di-Higgs events, the dependence of the di-Higgs production cross-section on \kappaLambda is parameterised as a function of the event rates for signals with $\kappaLambda = 1.0, 2.4, 3.0$. All other Higgs couplings are fixed to their SM values, and in particular no \kappaLambda dependence or uncertainties are assumed on the involved branching ratios. The resulting likelihood scans for the \kappaLambda parameter are shown in Figure~\ref{fig:result_kl_1D}. 
At 68\% confidence level, the expected precision on \kappaLambda ranges from 3.2\% in the case of considering only the statistical uncertainty, to 5.7\% for the systematic uncertainty scenario 3, as reported in Table~\ref{tab:syst_results}. This high precision is fully dominated by the \bbyy analysis, while the best precision reached by the \bbllMET analysis is roughly 20\% with statistical uncertainties only. 
Figure~\ref{fig:result_kl_kt_2D} shows as an additional interpretation the simultaneous constraints on \kappaLambda and the \kappaTop modifier of the Yukawa coupling between Higgs boson and top quark. Here only statistical uncertainties are considered. 

\section*{Acknowledgements} 
\noindent We thank Christophe Grojean for useful discussions, especially on the theory framework, and Michele Selvaggi for providing the \delphes detector scenario. This work is supported by the Deutsche Forschungsgemeinschaft (DFG, German Research Foundation) under Germany’s Excellence Strategy – EXC 2121 „Quantum Universe“ – 390833306.

\bibliographystyle{JHEP}
\footnotesize{\bibliography{bibl}}

\providecommand{\href}[2]{#2}\begingroup\raggedright\begin{thebibliography}{10}

\bibitem{ATLAS_HL_LHC_HH}
{\scshape ATLAS} collaboration,
\newblock \href{http://cds.cern.ch/record/2802127}{ATL-PHYS-PUB-2022-005}.

\bibitem{CDR}
{\scshape FCC} collaboration, \href{https://doi.org/10.1140/epjst/e2019-900087-0}{\emph{EPJST} {\bfseries 228} (2019) 755}.

\bibitem{fcc_dihiggs_paper}
M.L.~Mangano, G.~Ortona and M.~Selvaggi, \href{https://doi.org/10.1140/epjc/s10052-020-08595-3}{\emph{EPJC} {\bfseries 80} (2020) 1030}.

\bibitem{delphes}
J.~De~Favereau, C.~Delaere, P.~Demin, A.~Giammanco, V.~Lemaitre, A.~Mertens et~al., \href{https://doi.org/10.1007/jhep02(2014)057}{\emph{JHEP} {\bfseries 02} (2014) 057}.

\bibitem{powheg_one}
P.~Nason, \href{https://doi.org/10.1088/1126-6708/2004/11/040}{\emph{JHEP} {\bfseries 11} (2004) 040}.

\bibitem{powheg_two}
S.~Frixione, P.~Nason and C.~Oleari, \href{https://doi.org/10.1088/1126-6708/2007/11/070}{\emph{JHEP} {\bfseries 11} (2007) 070}.

\bibitem{powheg_three}
S.~Alioli, P.~Nason, C.~Oleari and E.~Re, \href{https://doi.org/10.1007/JHEP06(2010)043}{\emph{JHEP} {\bfseries 06} (2010) 043}.

\bibitem{hh_nnlo}
M.~Grazzini, G.~Heinrich, S.~Jones, S.~Kallweit, M.~Kerner, J.M.~Lindert et~al., \href{https://doi.org/10.1007/jhep05(2018)059}{\emph{JHEP} {\bfseries 05} (2018) 059}.

\bibitem{gghh_nnlo_kfactor}
S.~Amoroso, P.~Azzurri, J.~Bendavid, E.~Bothmann, D.~Britzger, H.~Brooks et~al., {\emph{arXiv} (2020) } [\href{https://arxiv.org/abs/2003.01700}{{\ttfamily 2003.01700}}].

\bibitem{madgraph_1}
J.~Alwall, R.~Frederix, S.~Frixione, V.~Hirschi, F.~Maltoni, O.~Mattelaer et~al., \href{https://doi.org/10.1007/JHEP07(2014)079}{\emph{JHEP} {\bfseries 07} (2014) 079}.

\bibitem{madgraph_2}
R.~Frederix, S.~Frixione, V.~Hirschi, D.~Pagani, H.-S.~Shao and M.~Zaro, \href{https://doi.org/10.1007/jhep07(2018)185}{\emph{JHEP} {\bfseries 07} (2018) 185}.

\bibitem{pythia}
T.~Sj{\"o}strand, S.~Mrenna and P.~Skands, \href{https://doi.org/10.1016/j.cpc.2008.01.036}{\emph{Computer Physics Communications} {\bfseries 178} (2008) 852}.

\bibitem{key4hep}
P.~Fernandez~Declara et~al., \href{https://doi.org/10.22323/1.398.0844}{\emph{PoS} {\bfseries EPS-HEP2021} (2022) 844}.

\bibitem{edm4hep}
``{\textit{EDM4hep data model}}.'' \url{https://edm4hep.web.cern.ch/}.

\bibitem{fccanalyses}
``{\textit{FCCAnalyses framework}}.'' \url{https://github.com/HEP-FCC/FCCAnalyses}.

\bibitem{keras}
A.~Gulli and S.~Pal, \emph{Deep learning with Keras}, Packt Publishing Ltd (2017).

\bibitem{tensorflow}
M.~Abadi et~al., \href{https://doi.org/10.48550/arXiv.1603.04467}{\emph{arXiv} (2016) } [\href{https://arxiv.org/abs/1603.04467}{{\ttfamily 1603.04467}}].

\bibitem{ATLAS_top_mass_dilep}
{ATLAS Collaboration}, \href{https://doi.org/10.1016/j.physletb.2016.08.042}{\emph{Physics Letters B} {\bfseries 761} (2016) 350}.

\bibitem{stranverse_mass}
C.G.~Lester, \href{https://doi.org/10.1007/jhep05(2011)076}{\emph{JHEP} {\bfseries 05} (2011) 076}.

\end{thebibliography}\endgroup

\end{document}